\documentstyle[12pt]{article}

\newcommand {\nn}    {\nonumber}
\newcommand {\vs}[1]  { \vspace*{#1 cm} }

\newcounter{eq}
\newcounter{sc}

\newcommand {\NP}   {Nucl.Phys.}
\newcommand {\PL}   {Phys.Lett.}
\newcommand {\PR}   {Phys.Rev.}

\def\overleftrightarrow#1{\vbox{\ialign{##\crcr
 $\leftrightarrow$\crcr\noalign{\kern-1pt\nointerlineskip}
 $\hfil\displaystyle{#1}\hfil$\crcr}}}

\newlength{\minitwocolumn}
\begin{document}


\begin{flushright}
EDO-EP-19\\
February, 1998\\
\end{flushright}
\vspace{30pt}

\pagestyle{empty}
\baselineskip15pt

\begin{center}
{\large\bf Path Integral Equivalence between Super D-string \\
and IIB Superstring
 \vskip 1mm
}

\vspace{20mm}

Ichiro Oda
          \footnote{
          E-mail address:\ ioda@edogawa-u.ac.jp
                  }
\\
\vspace{10mm}
          Edogawa University,
          474 Komaki, Nagareyama City, Chiba 270-01, JAPAN \\

\end{center}


\vspace{15mm}
\begin{abstract}
We show that the super D-string action is exactly equivalent 
to the IIB Green-Schwarz superstring action with some "theta
term" in terms of the path integral. Since the "theta term"
imposes the Gauss law constraint on the physical state but
contributes to neither the mass operator nor the constraints
associated with the kappa symmetry and the reparametrization,
this exact equivalence implies that the impossibility to 
disentangle the first and second class fermionic constraints 
covariantly in the super D-string action is generally inherited 
from the IIB Green-Schwarz superstring action except 
specific gauge choices which make the ground state massive, 
such as the static gauge. 
Moreover, it is shown that if the electric
field is quantized to be integers, the super D-string
action can be transformed to the IIB Green-Schwarz
superstring action with $SL(2, Z)$ covariant tension.

\vspace{15mm}

\end{abstract}

\newpage
\pagestyle{plain}
\pagenumbering{arabic}


\rm
\section{Introduction}

The recent discovery of an exact conformal field theory
description of Type II D p-branes makes it possible to
understand various non-perturbative properties of 
superstring theory \cite{Pol}. In particular, it is
remarkable that the Type IIB superstring in ten dimensions
has an infinite family of soliton and bound state strings 
permuted by $SL(2, Z)$ S duality group \cite{Schwarz}. 

More recently, several groups have presented supersymmetric
D-brane actions with local kappa symmetry \cite{Ceder, Aganagic,
Berg}, to which a covariant quantization has been performed 
by adopting the so-called static gauges $X^m = \sigma^m$ for 
the bosonic world volume reparametrization invariance and 
the covariant gauge $\theta_1 = 0$ for the fermionic kappa 
symmetry \cite{Aganagic}. In particular, since there is the 
$SL(2, Z)$ dual symmetry between IIB superstring and super
D-string as mentioned above, this means that a consistent 
covariant gauge fixing of the fundamental IIB superstring has 
been successfully achieved.
  
Soon after this success of the covariant quantization, a natural
question has been presented in the form that "Does it mean that 
the previous attempt to covariantly quantize the Green-Schwarz
string \cite{GS} missed the point, or something else happened? " 
\cite{Kallosh}   From the recent works \cite{Kallosh, Kamimura}, 
by now it seems to be generally accepted 
that the key ingredients for the covariant quantization of
the kappa symmetry can be summarized to the following two points.
One is that the consistent covariant gauge fixing of the kappa symmetry 
requires strictly massive ground state. For instance,
it is easy to see that the static gauge $X^1 = \sigma$ for the world 
sheet reparametrization symmetry in the case of super D-string 
\cite{Aganagic} makes the ground state massive.  This statement 
appears quite plausible from our past knowledges because the origin
of a difficulty of the covariant quantization exists in only
the massless sector of superstring. 

The other is that the space-time supersymmetry must be extended, 
i.e., $N > 1$, but this is only a necessary, not sufficient condition. 
To understand this importance, it is valuable to recall why we have not 
be able to quantize the Green-Schwarz superstring \cite{GS} 
in a Lorentz covariant way. 
The problem lies in the fact that in case of $N = 1$ supersymmetry 
it is impossible to disentangle covariantly 8 first class 
constraints generating the kappa symmetry and 8 second class 
constraints since the minimum off-shell dimension of covariant 
spinor (Majorana-Weyl spinor) in ten dimensions is equal
to 16. However, provided that there is $N = 2$ supersymmetry we have
a possibility of combining a pair of 8 first class constraints into
16 dimensional Lorentz covariant spinor representation, whose situation
precisely occurs in the procedure performed in \cite{Aganagic}. 
    
The purpose of this paper is two-fold. One of them is to show
the exact equivalence between the super D-string action and
the IIB Green-Schwarz superstring action with some "theta
term" in the framework of the path integral. 
It is valuable to point out that this action has been recently derived 
by the authors \cite{Kuriki} within the framework of the canonical 
formalism where they have also showed that canonical transformation 
to the type IIB theory with dynamical tension is constructed to 
establish the $SL(2, Z)$ covariance. By contrast, in this paper, 
we would like to show the following thing. Namely,
the "theta term" imposes the Gauss law constraint on the physical 
state but does not contribute to the mass operator and the
fermionic constraints, so 
this equivalence explicitly proves that the impossibility to 
disentangle the first and second class fermionic constraints 
covariantly in the super D-string action is generally 
inherited from the IIB Green-Schwarz superstring action 
except specific gauge choices 
which make the ground state massive, such as the static gauge. 
The other purpose is to show that if the electric
field is quantized to be integers, the super D-string
action can be transformed to the IIB Green-Schwarz
superstring action with $SL(2, Z)$ covariant tension without
appealing to any semiclassical approximation.

\section{ The IIB string with $SL(2, Z)$ covariant tension }

In this section, we consider the super D-string action in the flat 
space-time geometry, from which we would like to derive the
fundamental IIB superstring action with the $SL(2, Z)$ covariant
tension given by Schwarz formula \cite{Schwarz}. This derivation
is a straightforward generalization to the supersymmetric D-string
of the bosonic D-string performed in the reference \cite{de Alwis}
but it is worthwhile to expose the full detail of it since we
will make use of a similar technique in proving the exact equivalence 
between the super D-string and the IIB Green-Schwarz superstring
with some "theta term" in the next section.

The $\kappa$-symmetric super D-string action in the flat background
geometry is given by \cite{Aganagic}
\begin{eqnarray}
S = - n \int d^2 \sigma 
\left[ \ e^{- \phi} \left\{ \sqrt{- \det ( G_{\mu\nu} + 
{\cal F}_{\mu\nu} )} + \epsilon^{\mu\nu} \Omega_{\mu\nu} (\tau_1)
 \right\} + {\frac{1}{2}} \epsilon^{\mu\nu} \chi F_{\mu\nu} \right],
\label{2.1}
\end{eqnarray}
where 
\begin{eqnarray}
&{}& \det ( G_{\mu\nu} + {\cal F}_{\mu\nu} ) = \det G_{\mu\nu} + 
({\cal F}_{01})^2,
\ G_{\mu\nu} = \Pi_\mu^m \Pi_\nu^n \eta_{mn}, \nn\\
&{}& \Pi_\mu^m = \partial_\mu X^m - \bar{\theta}^A \Gamma^m 
\partial_\mu \theta^A,
\ {\cal F}_{01} = F_{01} - \epsilon^{\mu\nu} \Omega_{\mu\nu} 
(\tau_3), \nn\\
&{}& \Omega_{\mu\nu}(\tau_1) = \left\{ - \frac{1}{2} \bar{\theta}^A
\Gamma_m \tau_1 \partial_\mu \theta^A ( \partial_\nu X^m
- \frac{1}{2} \bar{\theta}^A \Gamma^m \partial_\nu \theta^A )
\right\} - (\mu \leftrightarrow \nu), \nn\\
&{}& \Omega_{\mu\nu}(\tau_3) = \left\{ - \frac{1}{2} \bar{\theta}^A
\Gamma_m \tau_3 \partial_\mu \theta^A ( \partial_\nu X^m
- \frac{1}{2} \bar{\theta}^A \Gamma^m \partial_\nu \theta^A )
\right\} - (\mu \leftrightarrow \nu).
\label{2.2}
\end{eqnarray}
Here $\mu, \nu, \cdots = 0, 1$ are the world sheet indices,
$m, n, \cdots = 0, 1, \cdots, 9$ ten-dimensional space-time ones, 
and $A = 1, 2$ is the two-dimensional spinor index.
Throughout this paper, we assume that the space-time metric takes 
the flat Minkowskian form defined as $\eta_{mn} = diag(- + + \cdots +)$.
Finally note that we confine ourselves to be only a constant dilaton
$\phi$ and a constant axion $\chi$, and set the antisymmetric
tensor fields to be zero.

Now we are ready to show how this super D-string action becomes 
a fundamental superstring action with the $SL(2, Z)$ covariant
tension by using the path integral. The equivalence in the case of the
bosonic string has already been shown in the paper \cite{Schmid,
de Alwis}. 
We shall follow the strategy found by de Alwis and Sato \cite{de Alwis}
since their method does not rely on any approximation. Incidentally,
it is necessary to use a saddle point approximation if we want to
apply this method to super D p-branes with $p > 1$ because of the 
nonlinear feature of the p-brane actions.

The major difference between super D-branes and super F-branes 
is the presence of $U(1)$ gauge field in the former. Hence
in order to show the path integral equivalence between two actions
it is enough to concentrate on the $U(1)$ gauge sector in the
super D-branes. Following de Alwis and Sato \cite{de Alwis}, 
let us define the theory in terms of the first-order Hamiltonian
form of the path integral. The canonical conjugate momenta 
$\pi_\mu$ corresponding to the gauge field $A_\mu$ are given by
\begin{eqnarray}
\pi_0 = 0, \ \pi_1 = \frac{n e^{-\phi} {\cal F}_{01}}
{\sqrt{-\det ( G_{\mu\nu} + {\cal F}_{\mu\nu} )}} 
- n \chi,
\label{2.3}
\end{eqnarray}
from which the Hamiltonian has the form 
\begin{eqnarray}
H = T_D \sqrt{- \det G_{\mu\nu}} + \epsilon^{\mu\nu} \Omega_{\mu\nu}
 (\tau_D) - A_0 \partial_1 \pi_1 + \partial_1 (A_0 \pi_1),
\label{2.4}
\end{eqnarray}
where $T_D$ and $\tau_D$ are defined as
\begin{eqnarray}
T_D &=& \sqrt{ ( \pi_1 + n \chi )^2 + n^2 e^{-2 \phi} }, \nn\\
\tau_D &=& ( \pi_1 + n \chi ) \tau_3 + n e^{- \phi} \tau_1.
\label{2.5}
\end{eqnarray}
 
Then the partition function is defined by the first-order Hamiltonian
form with respect to only the gauge field as follows:
\begin{eqnarray}
Z &=& \int {\cal D}\pi_1 {\cal D}A_0 {\cal D}A_1 \exp{ i \int 
d^2 \sigma ( \pi_1 \partial_0 A_1 - H ) } \nn\\
&=& \int {\cal D}\pi_1 {\cal D}A_0 {\cal D}A_1 \nn\\
& & {} \times \exp{ i \int 
d^2 \sigma \left[ - A_1 \partial_0 \pi_1 + A_0 \partial_1
\pi_1 - T_D \sqrt{- \det G_{\mu\nu}} 
- \epsilon^{\mu\nu} 
\Omega_{\mu\nu}(\tau_D) - \partial_1 ( A_0 \pi_1 ) \right] },
\label{2.6}
\end{eqnarray}
where we have canceled the gauge group volume against 
$\int {\cal D}\pi_0$. Note that if we take the
boundary conditions for $A_0$ and/or $\pi_1$ such that 
the last surface term in the exponential identically vanishes, 
then we can carry out the integrations over $A_\mu$, 
which gives us $\delta$ functions
\begin{eqnarray}
Z &=& \int {\cal D}\pi_1 \delta(\partial_0 \pi_1) 
\delta(\partial_1 \pi_1) \exp{ i \int d^2 \sigma 
\left[ - T_D \sqrt{- \det G_{\mu\nu}} 
- \epsilon^{\mu\nu} \Omega_{\mu\nu}(\tau_D) \right] }.
\label{2.7}
\end{eqnarray}
The existence of the $\delta$ functions reduces the integral 
over $\pi_1$ to the one over only its zero-modes. If we require
that one space component is compactified on a circle, these
zero-modes are quantized to be integers \cite{Witten}.
Consequently, the partition function becomes
\begin{eqnarray}
Z &=& \displaystyle{ \sum_{m \in {\bf Z}} } \exp{ i \int d^2 \sigma 
\left[ - t_D \sqrt{- \det G_{\mu\nu}} 
- \epsilon^{\mu\nu} \Omega_{\mu\nu}(\eta_D) \right] },
\label{2.8}
\end{eqnarray}
where 
\begin{eqnarray}
t_D &\equiv& \sqrt{ ( m + n \chi )^2 + n^2 e^{-2 \phi} }, \nn\\
\eta_D &\equiv& ( m + n \chi ) \tau_3 + n e^{- \phi} \tau_1.
\label{2.9}
\end{eqnarray}

To adapt the action in the above partition function to the form
of the Green-Schwarz superstring action \cite{GS}, one needs to 
replace $\eta_D$ in the argument of $\Omega_{\mu\nu}$ with $\tau_3$ by 
performing the $SO(2)$
rotation $\theta^A = u^{AB} \tilde{\theta^B}$ where $u$ is an
orthogonal matrix with constant elements. It is easy to
carry out this procedure by selecting the orthogonal matrix, for example,
\begin{eqnarray}
u = \frac{1}{b} \pmatrix{
m + n \chi + t_D & - n e^{-\phi}  \cr
n e^{-\phi}      & m + n \chi + t_D \cr },
\label{2.10}
\end{eqnarray}
with $b \equiv \sqrt{ ( m + n \chi + t_D )^2 + n^2 e^{-2 \phi} }$.
{}From the equation $u^T \eta_D u = t_D \tau_3$, we finally 
arrive at a desired form of the patition function
\begin{eqnarray}
Z &=& \displaystyle{ \sum_{m \in {\bf Z}} } \exp{ i \int d^2 \sigma 
\ t_D \left( - \sqrt{- \det G_{\mu\nu}} 
- \epsilon^{\mu\nu} \Omega_{\mu\nu}(\tau_3) \right) }.
\label{2.11}
\end{eqnarray}
{}From this expression of the partition function, we can read off
the action
\begin{eqnarray}
S = - t_D \left( \sqrt{- \det G_{\mu\nu}} 
+ \epsilon^{\mu\nu} \Omega_{\mu\nu}(\tau_3) \right),
\label{2.12}
\end{eqnarray}
which implies that the super D-string action is transformed to
the Type IIB Green-Schwarz superstring action with the
$SL(2, Z)$ covariant tension $t_D$. Note that we have obtained 
this result without making any approximation, which is a novel
feature of string theory.

\section{ The equivalence between super D-string and IIB 
superstring}

In what follows let us turn our attention to main 
purpose in this article, that is, to show the exact 
equivalence between the super D-string action and the 
IIB Green-Schwarz superstring action with some "theta term". 
One of the motivations behind this study is to clarify that 
both the super D-string action and the Green-Schwarz superstring action 
possess a similar structure with respect to the local symmetries,
which would in turn clarify the issue of the covariant quantization
of the kappa symmetry.

To this aim, one needs to sophisticate the machinery developed
in the previous section to adjust to the present problem.
In particular, one has to deal with not $\eta_D$ involving the
constant $m$ but $\tau_D$ including the field $\pi_1$. 
Moreover, a careful treatment of the functional measures
in the case at hand gives rise to an additional complication. 
Keeping these technical complications in mind, let us challenge
the above-mentioned problem.

As in the previous section, let us start with the super D-string
action (\ref{2.1}), and then define the partition function 
as in (\ref{2.6}). However, we should remark that the total partition
function $Z_T$ is really defined as
\begin{eqnarray}
Z_T = \int {\cal D}X^m {\cal D}\theta \ {\cal D}Y \ Z,
\label{3.1}
\end{eqnarray}
where ${\cal D}Y$ generically denotes the functional measures of ghosts, 
auxiliary fields e.t.c.  Of course, we can also consider the 
first-order Hamiltonian form of the path integral with respect to 
$X^m$ and $\theta$, but for simplicity here the second-order
Lagrangian form of the path integral is taken into account.
In order to rewrite the super D-string
action into the form of the Green-Schwarz superstring action,
first let us make the field redefinitions as follows:
\begin{eqnarray}
\tilde{X}^m = T_D^{\frac{1}{2}} \ X^m, \ \tilde{\theta} = 
T_D^{\frac{1}{4}} \ U^{-1} \ \theta,
\label{3.2}
\end{eqnarray}
where the orthogonal matrix $U$ is given by
\begin{eqnarray}
U = \frac{1}{B} \pmatrix{
\pi_1 + n \chi + T_D & - n e^{-\phi}  \cr
n e^{-\phi}      & \pi_1 + n \chi + T_D \cr },
\label{3.3}
\end{eqnarray}
with $B \equiv \sqrt{ ( \pi_1 + n \chi + T_D )^2 + n^2 e^{-2 \phi} }$.
The point to note here is that the functional measures in $Z_T$
(\ref{3.1}) are invariant under the field redefinitions (\ref{3.2}).
This is because if we fix the local symmetries the number of independent
degrees of freedoms associated with $X^m$ and $\theta$ is
respectively eight and sixteen so that the jacobian factors 
depending on $T_D$ exactly cancel out between bosons and fermions, 
and $\det U = 1$. 

We next move on to consider how various quantities in the partition
function (\ref{2.6}) change under (\ref{3.2}). For instance, we have
\begin{eqnarray}
\sqrt{- \det G_{\mu\nu}} &=& T_D^{-1} \sqrt{- \det \tilde{G}_{\mu\nu}}
+ f_1 ( \partial_\mu \pi_1 ), \nn\\
\Omega_{\mu\nu} (\tau_D) &=& \tilde{\Omega}_{\mu\nu} (\tau_3)
+ f_2 ( \partial_\mu \pi_1 ),
\label{3.4}
\end{eqnarray}
where $\tilde{G}$ and $\tilde{\Omega}$ are expressed in terms of
$\tilde{X}^m$ and $\tilde{\theta}$, and $f_1$ and $f_2$ are
certain functions whose concrete expressions are irrelevant for
the present arguments. Thus after the field redefinitions we obtain
the partition function
\begin{eqnarray}
Z &=& \int {\cal D}\pi_1 {\cal D}A_0 {\cal D}A_1 \nn\\
& & {} \times \exp{ i \int 
d^2 \sigma \left[ - A_1 \partial_0 \pi_1 + A_0 \partial_1
\pi_1 -  \sqrt{- \det \tilde{G}_{\mu\nu}} 
- \epsilon^{\mu\nu} \tilde{\Omega}_{\mu\nu}(\tau_3) + f ( \partial_\mu 
\pi_1 ) \right] } \nn\\
&=& \int {\cal D}\pi_1 {\cal D}A_0 {\cal D}A_1 \nn\\
& & {} \times \exp{ i \int 
d^2 \sigma \left[ \frac{1}{2} \pi_1 \epsilon^{\mu\nu} F_{\mu\nu}
-  \sqrt{- \det \tilde{G}_{\mu\nu}} 
- \epsilon^{\mu\nu} \tilde{\Omega}_{\mu\nu}(\tau_3) + f ( \partial_\mu 
\pi_1 ) \right] },
\label{3.5}
\end{eqnarray}
where we have rewritten the terms involving the gauge field
in terms of the field strength at the second stage.
The remaining problem is how to deal with the last term
$f ( \partial_\mu \pi_1 )$. Since this term is independent
of the gauge field $A_\mu$ we can absorb it into the first
term $\frac{1}{2} \pi_1 \epsilon^{\mu\nu} F_{\mu\nu}$
by performing an appropriate field redefinition of the gauge
field. Alternatively, if we allow to carry out the path
integral over the gauge field as in the previous section, 
we have the $\delta$-function $\delta(\partial_\mu \pi_1)$ 
so that the term $f ( \partial_\mu 
\pi_1 )$ vanishes identically.

After all, we reach the partition function which is exactly
equivalent to that of the super D-string
\begin{eqnarray}
Z = \int {\cal D}\pi_1 {\cal D}A_0 {\cal D}A_1
\exp{ i \int d^2 \sigma \left[ \frac{1}{2} \pi_1 \epsilon^{\mu\nu} 
F_{\mu\nu} -  \sqrt{- \det \tilde{G}_{\mu\nu}} 
- \epsilon^{\mu\nu} \tilde{\Omega}_{\mu\nu}(\tau_3) \right] }.
\label{3.6}
\end{eqnarray}
Even if the partition function was originally defined in terms of
the first-order Hamiltonian form with respect to the gauge
field, we can now regard it as the second-order Lagrangian
form of the path integral where $\pi_1$ must be viewed as 
an auxiliary field. From this viewpoint, the action is of the form
\begin{eqnarray}
S = - \int d^2 \sigma \left[  \sqrt{- \det \tilde{G}_{\mu\nu}} 
+ \epsilon^{\mu\nu} \tilde{\Omega}_{\mu\nu}(\tau_3) - \frac{1}{2} \pi_1 
\epsilon^{\mu\nu} F_{\mu\nu} \right] .
\label{3.7}
\end{eqnarray}
In this way we have derived  the action (\ref{3.7}) which is 
equivalent to the super-D string action even in the quantum level 
as well as the classical one if we regard $\pi_1$ as an auxiliary 
field. Note that the action (\ref{3.7}) has the form of the IIB 
Green-Schwarz action with the unit tension in addition to the 
"theta term" $\frac{1}{2} \pi_1 \epsilon^{\mu\nu} F_{\mu\nu}$. 
It is quite of interest to point out that the same action as 
(\ref{3.7}) has been recently derived by using the canonical 
transformations \cite{Kuriki}, in which it is shown that the 
constraints in the two actions (\ref{2.1}) and (\ref{3.7}) 
have one to one correspondence and two theories are canonically 
equivalent. 

Let us examine more closely what implication this "theta term"
has. First of all, if $\pi_1$ is quantized to be integers as 
investigated in the previous section, the "theta term" becomes 
the conventional two-dimensional theta term. Naively, when we 
neglect this true theta term it is obvious that we obtain 
the action (\ref{2.12}) with $t_D = 1$. Of course, 
this difference of the tension between two actions is 
inessential since we can change the overall value of the tension 
at will by the field redefinitions. 

Next, the more important point with respect to the "theta term"
in (\ref{3.7}) is that this term leads to the nontrivial constraint
on the physical state. To make the arguments clear let us consider
the canonical formalism. The canonical conjugate momenta to
the gauge field $A_\mu$ are given by
\begin{eqnarray}
\pi_0 = 0, \ \pi_1 = \frac{\delta S}{\delta \dot{A}_1}.
\label{3.8}
\end{eqnarray}
Here we have not treated $\pi_1$ as the dynamical variable, but
although we have done so we would obtain the same result through the
use of the Dirac bracket \cite{Dirac}. 
The consistency condition of the primary
constraint $\pi_0 \approx 0$ under time evolution gives rise to the 
Gauss law constraint $\partial_1 \pi_1 \approx 0$ as the secondary 
constraint. According to Dirac \cite{Dirac}, the first-class Gauss
law constraint must be imposed on the state as the physical state
condition
\begin{eqnarray}
\partial_1 \pi_1 |phys> = \partial_1 ( -i \frac{\delta}{\delta 
A_1} ) |phys> = 0.
\label{3.9}
\end{eqnarray}
The Gauss law constraint of two-dimensional gauge theory requires
that the physical states are of the form $\psi_p (A) = \exp{i p \int_
C \ A}$ \cite{Witten}. Since the gauge sector is 
completely decoupled from $X^m$ and $\theta$ sector in the 
action (\ref{3.7}), the total physical states are a direct product 
of $\psi_p (A)$ and the physical states of the $X^m$ and $\theta$ 
sector. Therefore the existence of the "theta term" in (\ref{3.7})
has no effect on the mass operator and the constraints' system in
the $X^m$ and $\theta$ sector. From this
reasoning, it is obvious that the super D-string action and 
the IIB Green-Schwarz superstring action share the common phase
structure except the $U(1)$ gauge sector.

\section{ Discussions }

In this paper, we have pursued the possibility of reformulating
the super D-string action in terms of the IIB Green-Schwarz 
superstring action. It has been shown that the super D-string
action is exactly equivalent to the IIB Green-Schwarz superstring
action with some "theta term". Because this "theta term" completely
decouples from the space-time coordinates $X^m$ and the spinor 
fields $\theta$ its existence does modify neither the mass spectrum
nor the common constraints' structure so that the infamous problem,
the impossibility of the covariant quantization of the kappa
symmetry still remains in both the actions as long as we do
not choose specific gauge conditions which make the ground state
massive. We believe that we have shed some light in this paper
on the relation betwen the super D-string and the fundamental
Green-Schwarz superstring.

Finally we would like to point out two issues for future work.
One issue is that in order to clarify the $SL(2, Z)$ duality
in more detail we should remove the restrictions on the flat
space-time, the vanishing antisymmetric tensor fields and
the constant background of the dilaton and the axion. It seems
to be interesting to apply the analyses done in this paper
to the more general super D-string and F-string. The other is
to understand how to realize the $SL(2, Z)$ duality in the IIB
matrix model \cite{IKKT}. We hope that we will return to
these issues in near future.

\vs 1
\begin{flushleft}
{\bf Acknowledgement}
\end{flushleft}
We are grateful to K. Kamimura, R. Kuriki, A. Sugamoto and
M. Tonin for valuable discussions. 
This work was supported in part by Grant-Aid for Scientific 
Research from Ministry of Education, Science and Culture 
No.09740212.

\vs 1

\end{document}